\documentclass[twocolumn,prl]{revtex4}
\usepackage{graphicx}
\usepackage{amsmath}

\newcommand{\be}{\begin{equation}}
\newcommand{\ee}{\end{equation}}
\newcommand{\beq}{\begin{eqnarray}}
\newcommand{\eeq}{\end{eqnarray}}

\def\nus{\mathrel{{\nu_s}}}

\def\nue{\mathrel{{\nu_e}}}
\def\numu{\mathrel{{\nu_\mu}}}

\def \lta {\mathrel{\vcenter{\hbox{$<$}\nointerlineskip\hbox{$\sim$}}}}
\def \gta {\mathrel{\vcenter{\hbox{$>$}\nointerlineskip\hbox{$\sim$}}}}

\def\t13{\mathrel{{\theta_{13}}}}
\def\y12{\mathrel{{\tan^2 \theta_{12}}}}
\def\c2{\mathrel{{\chi^2 }}}

\def\neff{\mathrel{{N_{eff}}}}

\def\yp{\mathrel{{Y_p}}}

\newcommand{\n}{neutrino}
\newcommand{\ns}{neutrinos}

\newcommand{\sn}{supernova}

\begin{document}
\title{Neutrinos, WMAP, and BBN}
\author{Lawrence M. Krauss$^{1,2}$}
\author{Cecilia Lunardini$^{2,3}$}
\author{Christel Smith$^2$}

\affiliation{$^1$School of Earth and Space Exploration, Arizona State University, Tempe AZ, 85287-1404}

\affiliation{$^2$Physics Department, Arizona State University, Tempe AZ, 85287-1404}

\affiliation{$^3$RIKEN BNL Research Center, Brookhaven National Laboratory, Upton, NY 11973}

\begin{abstract}

New data from WMAP have appeared, related to both the fractional energy density in relativistic species at decoupling and also the primordial helium abundance, at the same time as other independent observational estimates suggest a higher value of the latter than previously estimated.
All the data are consistent with the possibility that the effective number
of  relativistic species in the radiation gas at the time of Big Bang Nucleosynthesis may exceed the value of 3, as expected from a CP-symmetric population of the known \n\ species.
Here we explore the possibility that new \n\ physics accounts for such
an excess.  We explore different realizations, including \n\
asymmetry and new \n\ species, as well as their combination, and describe how existing constraints on neutrino physics would need to be relaxed as a result of the new data, as well as possible experimental tests of these possibilities.
%
% and study a previously unexplored combination of asymmetry and extra relativistic species
% the different mechanisms
%allowed by observation and experiment 
%whereby  new neutrino physics could account for such an excess, and
%examine their signatures at future experiments.  We also describe how
%existing constraints on neutrino physics now need to be relaxed as a
%result of the new data, and demonstrate a previously unexplored
%combination of effects that could increase neutrino densities during
%and after BBN. \comm{proofreading needed here..}
\end{abstract}

\maketitle

\section{Introduction}

One of the remarkable successes of Big Bang Nucleosynthesis (BBN) is the correct prediction of the overall magnitude of the light element abundances as a function of two fundamental parameters: the baryon density in the universe, and the number of light neutrino species.   Until recently BBN provided the only direct handles on these two fundamental parameters (i.e. \cite{kk93, Steigman:1977kc, Kang:1991xa}.   However with the discovery of primordial anisotropies in the Cosmic Microwave Background (CMB) radiation, the CMB has become a remarkable precision laboratory to constrain fundamental parameters in particle physics and cosmology.  It can now be used to test various ideas associated with BBN, and consistency checks can be applied  to probe new  physics.

It is therefore of some interest that recent results from both of these areas suggest the possibility the some new physics beyond the standard model may be at play.  Izotov and Thuan \cite{izotov} have recently analyzed the primordial helium abundance. For the primordial $^4$He mass fraction, $\yp$, they find  $Y_p = 0.2565 \pm 0.0010$(stat.)
$\pm  0.0050$(syst.), which is  higher, at the 2$\sigma$ level, than previous measurements  (see e.g. \cite{pdg} and references therein.)  Note that this value is in good agreement with another recent estimate \cite{skillman} although their quoted error is far smaller.  At the same time the new WMAP 7 year analysis \cite{Komatsu:2008hk} suggests both a high value of $\yp$ (with larger error bars), and independently a somewhat high value of the $N_{eff}$, the effective number of relativistic neutrino species present during last scattering ($N_{eff}= (\rho_{\rm rel} - \rho_\gamma) / \rho_{\nu_{\rm therm}}$, where $\rho_{\nu_{\rm therm}}=(7\pi^2/120)(4/11)^{4/3}T_\gamma^4$). Specifically, the WMAP 7 measurement is $N_{eff}=4.34+0.86-0.88$ \cite{Komatsu:2010fb}, about $1.4~\sigma$ higher than the standard  contribution of the known \n\ species, $\neff=3$. 

While the uncertainties in these estimates remain large, if confirmed experimentally in the future,  a scenario with high  $\yp$ and high $\neff$ 
%If these tentative suggestions are confirmed by later data this presents an 
would be an interesting challenge for particle physics. 
%An explanation that correlates the two excesses would be especially interesting, as well as non-trivial. Indeed, we are considering two facts that are very separated in cosmic time: an elevated helium abundance after BBN ($T \sim 0.2$ MeV) and extra energy density in relativistic degrees of freedom at matter-radiation decoupling ($T \sim 0.1$ eV). 
%
While multiple, uncorrelated, phenomena could explain it, both the Izotov and Thuan and the WMAP 7 results are clearly consistent with a single physical cause: that number of helicity degrees of freedom in the radiation gas at the time of BBN might have  been significantly higher than 3. The simplest and most conservative possibility for this involves new neutrino physics.
% It is, of course possible that the high $Y_p$ value and the CMB high $N_{eff}$ result could be unrelated.  But since both results are consistent with a single physical cause:, namely that the number of helicity degrees of freedom in the radiation gas at the time of BBN may be significantly higher than 3, this seems the most promising possible explanation of these effects, if indeed they are confirmed.  Similarly there are many possibilities for extra relativistic degrees of freedom involving as of yet undetected new light particles that may be populated by a variety of mechanisms in the early universe.  However, we consider perhaps the simplest and most conservative possibility here: namely that new neutrino physics might account for such effects.  \comm{editing needed here...improve flow}

Within the context of neutrino physics alone, two different possibilities arise.  One can consider mechanisms which affect both the energy density of neutrinos during BBN and the weak interaction rates that determine equilibrium nuclear abundances, such as would be  the case for a neutrino-antineutrino asymmetry. As a more minimal option, a correlation between two phenomena that are very separated in cosmic time--an elevated helium abundance after BBN ($T \sim 0.2$ MeV) and extra energy density in relativistic degrees of freedom at matter-radiation decoupling ($T \sim 0.1$ eV)--could have a common origin in an overabundance of weakly interacting and light (sub-eV) mass particles at the time of BBN.  During BBN this overabundance enhances the energy density, which, as we have alluded, in turn increases the expansion rate of the universe. This causes the weak reactions to freeze out earlier, resulting in a higher neutron-to-proton ratio and therefore a higher $Y_p$ \cite{Steigman:2010zz, Steigman:2010pa,Olive:1999ij, Steigman:2007xt, coulfac, Steigman:1977kc, Barger:2003rt, Barger:2003zg} .  Depending on their coupling to matter, the same particles could also affect $\yp$ by contributing to the weak reaction rates. At matter-radiation decoupling, they could still be relativistic and therefore contribute to $\neff$ as measured by WMAP7.
Other possibilities will also be briefly discussed. Previous strong constraints on both weak interaction physics and neutrino flavors need to be relaxed as a result of the current data, allowing possibilities that had previously been considered as ruled out. 

Clearly, suitable candidates should fit a number of conditions: (1) they should couple to matter and radiation strongly enough to be produced by thermal processes before BBN but (2) should not contribute too many extra degrees of freedom to the radiation gas, constrained in turn by measurements of $Y_p$.  Furthermore, (3) they are constrained by the existing cosmological bounds on the density of light extra degrees of freedom coming from a combination of data from the CMB, Large Scale Structure (LSS), Lyman Alpha Forest, and Baryon Acoustic Oscillations \cite{Komatsu:2010fb}: $\Omega_\nu  h^2< 0.006$ (95\% CL).  We consider specific scenarios and constraints in the following sections.
%
%There are also  model-dependent constraints on their production in laboratory and astrophysical sources (see e.g. \cite{.} for the case of sterile neutrinos). 
%
%\comm{I added the limit on $\Omega_\nu$ and added a mention to lab bounds (I removed it) and astrophysical bounds (e.g. SNe)}

%%%%%%%%%%%%%%%%%%%%%%%%%%%%%%%%%%%
\section{Neutrino Asymmetries and decays}
 
\begin{figure}
\includegraphics[width=3.5in]{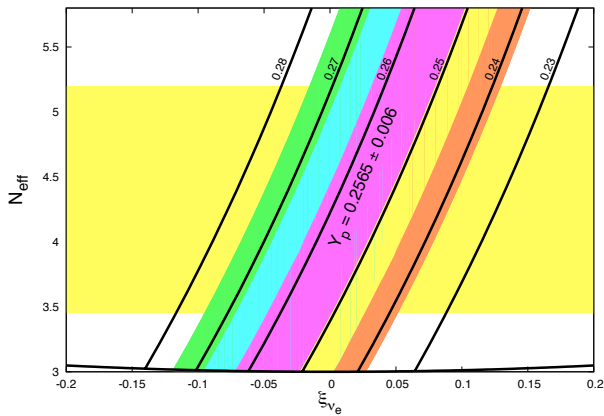}
\caption{$Y_p$ contours in the $\xi_{\nu_e}$ and $N_{eff}$ parameter space assuming neutrino flavor equilibration ($\xi_{\nu_e} = \xi_{\nu_\mu} = \xi_{\nu_\tau} $).  The horizontal light (yellow) band corresponds to the 1$\sigma$ WMAP 7 year result.  The black contours show a range of calculated values of $Y_p$ given model independent inputs of $\xi_{\nu_e}$ and $N_{\rm eff}$.  The shaded (colored) vertical bands mark the Izotov and Thuan 1$\sigma$, 2$\sigma$, and 3$\sigma$ ranges of $Y_p$.  The bottom black curve shows the contribution to $N_{eff}$ from neutrino asymmetries alone.}
\label{Yp}
\end{figure}

An overabundance of neutrinos with respect to anti-neutrinos or vice-versa,  $L_\nu \equiv (n_\nu - n_{\bar \nu})/n_\gamma $, is defined by a non-zero degeneracy parameter, $\xi$:  $L_\nu =\pi^2/(12 \zeta(3))(T_\nu/T)^3(\xi + \xi^3/\pi^2)$.  The total change in the effective number of relativistic species resulting from asymmetries in each flavor, $\xi_{\nu_\alpha}$, is given by
\begin{equation}
\Delta N_{eff} = \sum_{\alpha=e,\mu,\tau} \left[ {{30}\over{7}} \left({{\xi_{\nu_\alpha}}\over{\pi}}\right)^2 + {{15}\over{7}}\left({{\xi_{\nu_\alpha}}\over{\pi}}\right)^4\right].
\label{deltaN}
\end{equation}
In most theoretical scenarios, lepton and baryon asymmetries are enforced
to be of the same order by sphalerons \cite{Kuzmin:1985mm}, so that $L_\nu \sim 10^{-10} - 10^{-9}$. However, several scenarios have been proposed in which a large lepton asymmetry can be generated while preserving a small baryon asymmetry, using e.g., GUT models, the Affleck-
Dine mechanism, Q-balls, resonant oscillations, etc. \cite{Harvey:1981cu,Casas:1997gx,MarchRussell:1999ig,McDonald:1999in,Kawasaki:2002hq,Gu:2010dg}. Therefore, here we assume $L_\nu$ as independent from the baryon asymmetry and consider only direct constraints on it from \n\ physics. 

While asymmetries in all flavors contribute to an increase in energy density,  only an asymmetry in the electron flavor influences the weak neutron-proton interconversion processes.  For this reason, the sensitivity of BBN to $\xi_{\nu_e}$ is remarkably high: $|\xi_{\nu_e}| \lta few~ 10^{-2}$ is needed for compatibility with measured abundances (see e.g.\cite{Serpico:2005bc, Kang:1991xa, abfw, Kneller:2004jz, Simha:2008mt, Kneller:2001cd, Barger:2003rt, Barger:2003zg}).  This applies also to the asymmetries in the other flavors at the time BBN, since  oscillations should produce an at least approximate flavor equilibration before BBN \cite{Lunardini:2000fy, wong, abb, dolgov}.

Under such strong constraint,  
%%
%Neutrino oscillations in the early universe may begin early enough to equilibrate asymmetries among flavors, implying that the three neutrino degeneracy parameters must be approximately equal \cite{Lunardini:2000fy, wong, abb, dolgov}.  If this is the case, the BBN bound on the electron asymmetry, $-0.04 \leq \xi_{\nu_e} \leq 0.07$ \cite{Serpico:2005bc}, will apply to all flavors, although this quoted  bound will be relaxed both by the possibility of higher $Y_p$ and $\neff $.  Even with a somewhat looser bound on $\xi_{\nu_e}$,  the condition that the asymmetries in each flavor must be nearly equal would imply that 
%
neutrino asymmetries alone generally cannot account for a $\Delta N_{eff} \sim 1$.  An interesting exception is the somewhat fine-tuned scenario of initial (pre-equilibration) flavor asymmetries that are large and opposite in sign.  After equilibration, a surviving $\Delta N_{\rm eff} \sim 1$ can be realized, together with sufficiently small asymmetries that satisfy BBN bounds \cite{Pastor:2008ti}. 
This reopens the possibility of having, at BBN, virtually any combination of $\xi_{\nu_\alpha}$ and energy density.   In general, asymmetries could coexist with other effects (e.g., a sterile neutrino, see next section) that could independently increase $\neff$. Therefore an analysis that treats asymmetries and energy density as independent is necessary to find the most general constraints on both. 

Here we perform such a study,  using  a modified version of the Kawano/Wagoner BBN code described in detail in Ref.~\cite{ourcode,coulfac}.  In Fig. 1 we illustrate the interplay between asymmetries and $\neff$ by plotting the $Y_p$ abundance yield isocontours in the $N_{eff}$ - $\xi_{\nu_e}$ parameter space.  This figure shows calculations for model-independent inputs of $N_{eff}$ over a wide range of neutrino asymmetries, where we have adopted the condition of neutrino equilibration of asymmetries, so that the allowed range of asymmetries is small, and the direct effect of such asymmetries on $\neff$ is minimal (as displayed in the lower curve, which shows the extra direct contribution to $\neff$ from such asymmetries).  The horizontal band for $\neff$ corresponds to the 1$\sigma$ WMAP 7 year result quoted earlier.  

The BBN code used to make Fig. 1 differs from others in a number of ways, mostly in the treatment of the weak processes.  It allows for calculations that include both the effects from higher relativistic degrees of freedom and neutrino asymmetries, which is not the case for the Kawano/Wagoner code.  This code gives a standard BBN $Y_p$ yield that is $\sim 0.004$ lower than other calculations due to the full numerical integration of each weak reaction rate.  This figure is not intended to provide new constraints on, or a best fit for, neutrino asymmetries and/or $N_{eff}$.  It is simply a model-independent tool to illustrate the fact that a wider range of allowed $N_{eff}$ parameter space loosens the BBN limit on the neutrino asymmetries or lepton numbers.

% If the equilibrated asymmetry condition is not present however, then one can choose any combination of asymmetries to dial any value of $N_{eff}$ and $Y_p$.  \comm{I only meant here that if you didn't need to satisfy equilibration, then you could have any arbitrarily high asymmetry in the mu and tau flavor to give any value of $N_{eff}$ and then choose any value of electron asymmetry to provide the correct helium abundance. -cs} 
%
%
%It is therefore interesting to consider the interplay of $N_{eff}$ with neutrino asymmetries for a general scenario in which these two effects are largely independent.   

%
This figure demonstrates explicitly the (known) fact that  the neutrino electron degeneracy parameter has significant leverage on $Y_p$.  Positive $\xi_\nu$ drives the neutron destruction process, $\nu_e+n\rightleftharpoons p+e^-$ forward and and Pauli-blocks the reverse neutron production reaction resulting in a lower $Y_p$.  Negative $\xi_\nu$, corresponding to an overabundance of anti-neutrinos, drives anti-neutrino capture forward  $\bar\nu_e+p\rightleftharpoons n+e^+$ and suppresses the reverse proton production reaction resulting in more helium.   This figure displays explicitly the novel recognition that to obtain a given isocontour for $Y_p$, one can increase $\xi_\nu$ while also increasing $N_{eff}$, and it provides a quantitative estimate of the interplay between these effects.  This interplay has not been directly examined quantitatively before, although a correlation between the two parameters was noted in a fit to BBN data  \cite{Simha:2008mt} and in an analysis of the baryon-to-photon ratio\cite{Barger:2003rt}. 

The figure also shows how the constraints on $\xi_{\nu_e}$ change with a change of $\neff$ and can thus be relaxed compared to previous limits. For example, if we allow $\neff$ in the $1 \sigma$ WMAP7 interval and $\yp$ in the $3\sigma $ interval of  Izotov and Thuan we get an approximate allowed range of
\begin{equation} 
-0.14 \lta \xi_{\nu_e} \lta 0.12, 
\end{equation}
larger than the usually quoted constraint  $-0.04 \leq \xi_{\nu_e} \leq 0.07$ \cite{Serpico:2005bc}. 
%
% (Note that this figure is only relevant for interacting neutrinos, and not sterile neutrinos because it has been shown that asymmetries suppress their thermalization in the early universe \cite{FV,fv97,Cirelli:2004cz}.\comm{are we saying this multiple times?}) 
 %
 While these results present a model-independent exploration of parameter space, with each point involving a full BBN code calculation of $Y_p$ for a given value of $N_{eff}$ and $\xi$, whether any specific point in parameter space is actually realizable however, will depend upon specific model building issues. 

We conclude this section by briefly mentioning another possibility for adding 
extra relativistic energy density both during BBN times and during the matter-radiation equality epoch: extra particles that are unstable. To produce higher $Y_p$, these particles must contribute to the relativistic energy density during the \lq\lq weak freeze out\rq\rq\ period in BBN, implying masses $m \leq 1$ MeV.  They would then be required to decay by the time of matter-radiation equality ($T\sim 1$\ eV), where the CMB measurements infer extra relativistic degrees of freedom (see for example \cite{Cuoco:2005qr,Serpico:2004nm}). 

However, additional cosmological constraints imply that the physics of such particles is so finely tuned as to be implausible.  Their decay can only be into neutrinos so as not to produce high-energy photons which result in subsequent deuterium photo-disassociation \cite{Holtmann:1998gd}.  Furthermore, the additional particles must decay quickly enough so that they don't subsequently dominate the energy density of the universe once the temperature falls below their mass.  The dual requirements of being primarily weakly interacting and also decaying within the appropriate time window are extremely difficult to satisfy. 

%%%%%%%%%%%%%%%%%%%%%%%%%%%%%%%%%%%
\section{Sterile and Right handed Neutrinos}

The minimal scenario to explain both high $\neff$ and high $\yp$ with \n\ physics is a  light sterile \n \ (see also \cite{Hamann:2010bk}).  
Indeed, sterile \ns\ easily fit the conditions we have outlined above: by definition they are weakly interacting, they can be produced before BBN, and they are allowed, by laboratory and astrophysical bounds, to be of sub-eV mass. 

Specifically, if the neutrino masses come from a See-Saw-like mechanism, active-sterile oscillations arise naturally due to the mixing of active \ns\ with the charge-conjugate of one or more right handed neutrinos. Depending on the mixing and masses, the interplay of oscillations and collisions can populate the sterile \ns\ before BBN (see e.g., \cite{Smirnov:2006bu} and references therein). 
While in minimal See-Saw models the sterile \ns\ are too heavy to contribute to $\neff$, several non-minimal scenarios (e.g.,  \cite{deGouvea:2005er,deGouvea:2006gz})  include sterile \ns\ lighter than an eV. 
%
%Depending on their masses and mixings with the active \ns, the sterile \ns\ can be produced abundantly  prior to BBN via active-sterile oscillations.   They would then remain relativistic through the CMB epoch, and survive to this day, influencing large scale structure formation like the active \ns\ do. 

\begin{figure}[htbp]
%\begin{figure}[t]
  \centering
% \title{Fig. 6}
% \newline
% \newline
 \includegraphics[width=0.45\textwidth]{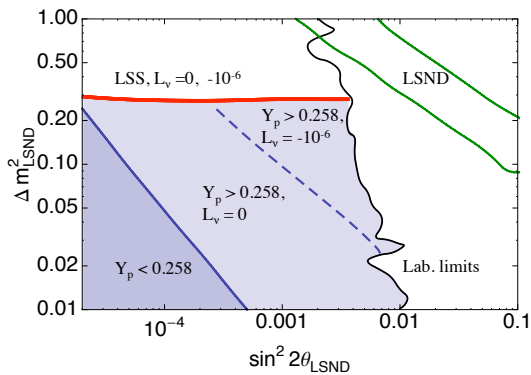}
 \caption{Limits on the sterile \n\ oscillation parameters $\Delta m^2_{LSND}$ (in ${\rm eV^2}$) and $\theta_{LSND}$ from several \n\ experiments  and from the cosmological bound $\Omega_\nu h^2< 0.006$.  The shaded regions are allowed. For zero asymmetry, the dark (light) shaded area corresponds to $\yp<0.258$ ($\yp>0.258$). The dashed line represents the boundary of  the  $\yp>0.258$  region for asymmetry $L_\nu= -10^{-6}$.  
The region favored by the LSND data, now almost entirely excluded, is shown as well.  See text for details.  }
\label{sterileplot}
\end{figure}
Detailed studies exist on the effect of one sterile \n\ ($\nus$ from here on) on $\yp$ and $\neff$ (e.g. \cite{Cirelli:2004cz,Chu:2006ua,Hamann:2010bk}). Here we consider one such case, assuming a hierarchical spectrum with the predominantly sterile state being the most massive, $\Delta m^2_{41} \gg |\Delta m^2_{31}| $.    Fig. \ref{sterileplot} (adapted from the results of \cite{Chu:2006ua}) refers to the specific case in which $\nus$ mixes with $\numu$ and $\nue$, as needed to interpret the LSND anomaly \cite{Aguilar:2001ty}.   It is however representative of  the general situation. The four-\n\ mixing scheme can be reduced to an effective two-\n\ mixing with parameters $\Delta m^2_{LSND} \simeq \Delta m^2_{41} =m^2_4 - m^2_1$ and $\theta_{LSND} \simeq  \theta_{es}\theta_{\mu s}$ \cite{Cirelli:2004cz} (Figure \ref{sterileplot} refers to the case $\theta^{1/2}_{LSND} \simeq  \theta_{es} \simeq \theta_{\mu s}$, which is conservative in that it generally corresponds to minimizing the high $\yp$ region, see \cite{Chu:2006ua}.).

We consider both sterile neutrinos with and without lepton asymmetry, and we comment on the case of zero asymmetry first.  In the plot we highlight the region of interest: the area (light shaded in the plot) where $\nus$ is produced abundantly prior to BBN, thus causing a high $\yp$ via its contribution to the energy density, and contributing to $\neff$ at matter radiation-decoupling. Specifically, the region corresponds to  $Y_p \geq 0.258$ and $\neff = 3.8 - 4$, i.e., a nearly or completely populated sterile state.  It is bounded from below by the ``thermalization line", where the $\nu_s$ production rate is comparable to the cosmic expansion rate \cite{Enqvist:1991qj}. and spans more than an order of magnitude in each parameter, extending down to $\sin^2 2\theta_{LSND}\sim 10^{-5}$. 
The region is constrained in mixing by several terrestrial experiments (mainly Karmen, Bugey, SuperK, CDHS \cite{cdhs,Declais:1994su,Armbruster:2002mp,Fukuda:2000np} \footnote{The recent MiniBooNE results \cite{AguilarArevalo:2010wv} are not yet competitive with older limits, due to low statistics.}  and in $\Delta m^2_{LSND}$  by the cosmological bound on $\Omega_\nu$.  
For $\neff=4$, this bound gives:
\beq
 \Delta m^2_{41} \lta 0.28~{\rm eV^2}   \hskip 0.5 truecm m_4 \lta 0.53~{\rm eV}~,
\label{masslimit}
\eeq
assuming, conservatively, $m_1 \simeq m_2 \ll m_3 \simeq 0.05 $ eV, as given by oscillation data for the normal mass hierarchy \cite{pdg}. 
%
%The main element of the figure is the region of the  $\Delta m^2_{LSND}-\sin^2 2 \theta_{LSND}$ plane corresponding to  $Y_p \geq 0.258$.  For parameters in this region, the sterile \n\ is produced abundantly prior to BBN, increasing $\neff$ to about $\neff \simeq ...$ (see fig. ... ), which means a nearly thermalized sterile \n\ population.  The higher $\neff$ is then responsible for the increased $\yp$.   
%
% The region of high $\yp$ is constrained in $\Delta m^2_{LSND}$ by LSS bound on the sum of all \n\ masses, and by data from terrestrial experiments, which are more sensitive to the mixing.  
%We see that the region of high $\yp$ is wide, spanning more than an order of magnitude in each parameter. 
We stress that the area we are considering was  interpreted as excluded by BBN until recently. Our new perspective reopens this possibility.  %We also observe how the improved cosmological bound on $\Omega_\nu$ 

Results similar to those in fig. \ref{sterileplot} are obtained for other active-sterile mixing scenarios, such as those in which $\nu_s$ mixes with one active flavor or with one \n\ mass eigenstate.  The mixing of $\nu_s$ with $\nu_3$  is the least constrained  because of  the strong constraint on the  $\nue$ component of $\nu_3$ \cite{Cirelli:2004cz}.  
%Interestingly, $\nu_s- \nu_3$ mixing with exotic \n-matter interaction has been considered recently \cite{.} as a way to explain a hint of apparent CP violation at MINOS \cite{.} in the form of difference between the \n\ and antineutrino channels.  \comm{I cut the statement about the parameters of the model being in the high $\yp$ region. This is because the model relies on exotic interactions and so we can't compare it with a high $\yp$ region calculated with standard physics.}
%
% The active-sterile oscillation parameters favored by the MINOS data are .... , which lie in the high $\yp$ region, as shown explicitly in \cite{.}.
%Thus, if the apparent CP violation at MINOS becomes statistically significant, the case for a light sterile \n\ producing high $\neff$ and  high $\yp$ in the early universe would be reinforced. 

In the presence of lepton asymmetry, the production of sterile \ns\ via oscillations is suppressed \cite{Dolgov:2003sg,FV,Foot:1996qc,Chu:2006ua}.  This effect is due to the term in the potential describing \n-\n\ forward scattering, which suppresses the active-sterile mixing for \ns\ or antineutrinos and is zero for a symmetric \n\ population.  
Fig. \ref{sterileplot} shows how the high $\yp$ region changes with the increase of $L_\nu$ which is assumed to be equal for all flavors. 
As a result of the suppression, with lepton asymmetry the region of high $\yp$ is reduced to a smaller area at high $\Delta m^2_{LSND}$ (fig. \ref{sterileplot}) and eventually disappears for $L_\nu \simeq 10^{-5} $ \cite{Chu:2006ua} as the mass required to populate the sterile \n\ becomes excluded by the bound on $\Omega_\nu$.  Thus, a single light sterile neutrino can give only a subset of the region in our fig. \ref{Yp}. This subset has $3\leq \neff\ \leq 4$, with $\neff\simeq 4$ being realized only for $\xi_{\nue} \simeq 0$.
 
 %It is also worth noting that the improved bounds on $m_{\nu}$ from WMAP7 has effectively removed the remaining phase space appropriate to explaining LSND, even in the presence of asymmetries. 

The conclusions on the suppression of the production of $\nu_s$ are generally true for a wide range of lepton asymmetry, $L_\nu \sim 10^{-5} - 1$, provided that the asymmetry is constant over the characteristic time scale of the sterile \n\ production. For $L_\nu \sim 0.1 - 1$ the presence of the sterile state can modify $\yp$ through a modification of the spectra of the active states, while not affecting $\neff$ \cite{abfw, kfs, sfka}. Therefore we do not consider this scenario here.  
%
%\comm{Here is some text that refers to the Volkas paper (and others: Shi 1996, Dolgov and Villante, Volkas and Thomson, ... )} 
%
A more diverse phenomenology is expected if the asymmetry varies over the time of $\nu_s$ generation: active-sterile oscillations can actually generate an asymmetry in the active flavors that can survive and affect the weak reaction rates \cite{Foot:1995qk,Shi:1996ic}. Although an updated analysis on this is not available, from existing  studies \cite{Foot:1996qc} we infer that if the sterile \n\ contributes substantially to $\neff$, and therefore its mass is below the LSS bound, the generated asymmetry is $L_\nu \lta 10^{-2}$,  not sufficient to impact $\yp$ via BBN reaction rates.  Therefore, this effectively reduces to the case $L_\nu=0$.

If the \n\ mass arises from a Dirac mass term only, the right handed neutrino(s) associated to it are not produced via oscillations from the active states and therefore do not play the role of sterile \ns\ as described above.  Still, a number of models exist  in which right handed \ns\ are of sub-eV mass, and couple to the Standard Model particles strongly enough to be populated substantially prior to BBN \cite{Steigman:1979xp,Olive:1999ij,Barger:2003zh}.  
% An example is models with an extra $U(1)^\prime$ symmetry, where the right handed \ns\ are produced in the early universe by their coupling with the new, TeV scale $Z^\prime$ boson \cite{.}.  
%
 A detailed analysis  in the context of an $E_6$ symmetry \cite{Barger:2003zh} shows how, indeed, increased $N_{eff}$ and  $Y_p$ are expected due to the right handed \n\ production.  Compatibility with BBN translates into lower limits on the mass of the $Z^\prime$.  These limits will be relaxed for increased $N_{eff}$ and  $Y_p$; the new expected mass range will be close to or overlap with the limits from SN1987A \cite{Barbieri:1988av}, and therefore measurements from a future galactic \sn\ could test such class of models, as we describe below.

\section{Experimental Signatures}

A sterile \n\ with parameters in our high $\yp$ region   has a  number of implications for future detectors.   

Beta decay and neutrino-less double beta decay experiments, designed to  measure the \n\ mass, would probe a fourth light mass state that mixes with the electron neutrino (see e.g. \cite{Goswami:2005ng}).
The next generation of reactor \n\ experiments 
%like Double CHOOZ and Daya Bay \cite{.} 
%
will probe this region beyond the existing limits, to an extent that depends on the specific model and on $\theta_{13}$ \cite{deGouvea:2008qk}.  Neutrino beams will also allow to search for sterile states, probing different parameters depending on their energy and baseline \cite{Dighe:2007uf}. For a $\sim 10$ GeV beam and $\Delta m^2_{41} \lta 0.1~{\rm eV^2}$, a baseline of $L \sim 2 \pi E/\Delta m^2_{41} \gta 100 $ Km is required.  Signatures of  a sterile state would be disappearance of the active flavors and anomalous differences in the oscillation pattern of \ns\ and antineutrinos due to refraction in the Earth. A recent example of the latter involves a sterile \n\ with parameters in our  region of interest (high $\yp$ and high $\neff$) \cite{Engelhardt:2010dx}, and is still interesting in the light of the hint of neutrino-antineutrino differences at MINOS \cite{AguilarArevalo:2010wv}. 

It was observed \cite{Chu:2006ua} that the suppression produced by an asymmetry relaxes the cosmological bound on $\Delta m^2_{LSND}$, thus allowing parameters that explain the LSND anomaly.
% These however correspond to low $\yp$ and $\neff \simeq 3$; therefore a future measurement of high $\yp$ and high $\neff$ due to the existence of a sub-eV sterile neutrino could exclude LSND. 
As fig. \ref{sterileplot} clarifies, however, these parameters correspond to low $\yp$ and $\neff \simeq 3$ (the sterile state is not populated).  Therefore, if  MiniBOONE confirms LSND and high $\neff$ and $\yp$ are established, less minimal scenarios, beyond a CP-symmetric system of four \ns, would have to be considered. 
%
%this would reveal a system of five \ns, two of which sterile. \comm{ I don't think this statement is accurate because you need an asymmetry to compensate for the LSND mass neutrino, but this asymmetry will also suppress the production of our sub-eV neutrino...unless you have other new physics, such as a low reheat temperature for inflation (gelmini-2004) or perhaps even mass-varying neutrinos. -cs}

Important astrophysical  tests of a light sterile \n\ would come from 
atmospheric \ns\ and 
a future \sn\ \n\ detection \footnote{Bounds from  solar \ns\ refer to lower sterile masses, $\Delta m^2_s \lta 10^{-4}~{\rm eV^2}$, so they do not apply to the hierarchical mass scenario that we discuss here. } 
 Higher precision measurements of atmospheric \ns\ would extend the currently probed region of parameters. 
 % If the hint of low $\mu/e$ ratio in the data, is confirmed, a sterile \n\ with  $\Delta m^2 \simeq ... $ would be favored \cite{.}. \comm{check the parameters and see if they are relevant for us... see Peres-Smirnov paper}. 
 The detection of 0.1-1 TeV atmospheric \ns\ at IceCUBE would facilitate searches of 
 sterile \ns\ in the higher $\Delta m^2 $ range of our region, for which the active-sterile mixing is enhanced by matter effects \cite{Nunokawa:2003ep,Choubey:2007ji}.  

In a supernova, a sterile \n\ in our high $\yp$ region could be produced via resonant oscillations and cause a suppression  of the active \n\ signal.  Due to partial violation of adiabaticity \cite{Cirelli:2004cz,Smirnov:2006bu}, the suppression would be moderate, at the level of tens of per cent. It will be detectable with a future galactic \sn\ if precise theoretical predictions of supernova \n\ emission are available. 
%\comm{see Smirnov-Zukanovich... note the definition of our LSND angle.. it is the product of two angles, so it is not easy to compare with Cirelli... }

Compared to a sterile \n, a right handed \n\ would be more difficult to test experimentally, because it requires non-oscillation tests.  
 A possibility is  to look  for the new gauge bosons that couple to the right handed states, at man made or cosmic accelerators.  Among the latter, the \n\ burst from a galactic \sn\ would be sensitive to the the production of right handed states via their effect on the cooling rate, as mentioned previously. 
% 
% Due to such couplings, the production of right handed states affects the  cooling rate of a \sn\ \cite{}, and thus can be tested with a  a future galactic 
 %
%  (and therefore its observed \n\ burst) depending on their coupling and mass. 
Constraints on a right handed \n\ will be highly model-dependent. 

If a high $\neff$ and high $\yp$ are confirmed and other data exclude extra \n\ species that can be populated before BBN, one would have to consider more finetuned scenarios like large and opposite \n\ asymmetries in the different flavors.  This would indicate CP violation in \ns, with profound implications on our understanding of 
the lepton sector. 
%
%although  ....... 
%
 While to uniquely trace back to a model that generates opposite asymmetries might not be possible,  a large number of mechanisms that predict more natural, comparable, asymmetries would be disfavored. 
Note, however, that it would be very difficult to confirm this scenario independently, because direct CP violation experiments in the neutrino sector would not probe the CP violating effects of relevance during BBN.  Therefore precision cosmology alone might be required to more firmly constrain the neutrino sector.   

Precision cosmological tests may be possible in the not-too-distant future.  Direct kinematic constraints on neutrino masses from large scale structure tests, in particular from probes of galaxy clustering via the Sunyaev-Zeldovich effect on CMB, as currently being explored in the South Pole telescope \cite{spt1,spt2} could improve existing mass constraints by a factor of 2-5.   At the same time, the Planck satellite will improve constraints on $N_{eff}$ by increased sensitivity to high-$l$ CMB anisotropies.   Finally, direct measurements of primordial $Y_p$ might be possible in the more distant future if spectral sensitivity to the CMB can be improved by several orders of magnitude. 

To conclude, if the tentative new results on $\neff$ and $\yp$ are confirmed, they open up new possibilities for neutrino physics which may be accessible in the future by ground-based experiments and astrophysical probes.  Specifically we find:

\begin{itemize}

\item{A new relaxed constraint on possible electron neutrino asymmetry $-0.14 \lta \xi_{\nu_e} \lta 0.12,$}

\item{if $ \neff \approx 4$, a bound $\Delta m^2_{41} \lta 0.28~{\rm eV^2} ,  \hskip 0.5 truecm m_4 \lta 0.53~{\rm eV}~$ on sterile neutrino masses under fairly conservative assumptions about a mass hierarchy }

\item{a new quantitative relation between the effects of possible independent variations in $\xi_{\nu_e}$ and $ \neff$ on $\yp$.}

\item{a new set of possible astrophysical and experimental signatures that might further probe these scenarios}

\end{itemize}

%If the CMB and BBN results hold up, the new neutrino physics possibilities we have mentioned can result in interesting experimental signatures, which we briefly conclude with by reviewing. 

%-asymmetry, CP violation:  $\theta13$ experiments

%-sterile neutrinos: MINOS etc?  \comm{anything we can say on the neutrino-antineutrino anomalies of minos and miniboone?}

%\comm{new atmospheric expts? Future SN observations? decay of sterile into active + photon? minos searches for sterile neutrinos? }
\begin{acknowledgments}
We would like to acknowledge helpful discussions with  George Fuller, Chad Kishimoto, Gary Steigman, and Francesco Villante.  LMK acknowledges support from the DOE for this work, and 
C.L. and C.S. acknowledge the support of the NSF under Grant No. PHY-0854827. 
\end{acknowledgments}

\bibliography{mybiblio}
%\begin{thebibliography}{99}
%\bibitem{xxx} xxx, Phys. Lett. {\bf xxx}, 229 (1992).
%\end{thebibliography}
\end{document}